\documentstyle[prb,multicol,aps,epsf]{revtex}
\input epsf.sty
\begin{document}
\title{Concept of local polaritons and optical properties of mixed polar crystals\footnote{Submitted to PRB}}
\author{ Lev I. Deych$^{\dagger}$ Alexey Yamilov,$^{\ddagger}$ and Alexander A. Lisyansky$^{\ddagger}$}

\address{$^{\dagger}$Department of Physics, Seton Hall University, 400 South Orange
Ave, \\ South Orange, NJ 07079}
\address{$^{\ddagger}$Department of Physics, Queens College of CUNY, Flushing, NY 11367}
\date{\today}
\maketitle

\begin{abstract}
The concept of local polaritons is used to describe optical properties of mixed crystals in the frequency region of their {\it restrahlen} band. It is shown that this concept allows for a physically transparent explanation of the presence of weak features in the spectra of so called one-mode crystals, and for one-two mode behavior. The previous models were able to explain these features only with the use of many fitting parameters. We show that under certain conditions new impurity-induced polariton modes may arise within the {\it restrahlen} of the host crystals, and study their dispersion laws and density of states. Particularly, we find that the group velocity of these excitations is proportional to the concentration of the impurities and can be thousands of times smaller then the speed of light in vacuum.   
\end{abstract}



\section{Introduction}

The optical properties of mixed polar crystals have been attracting a great deal of interest since the 1950's. Main efforts have been directed to experimental and theoretical studies of the concentration dependencies of fundamental transverse  (TO) and longitudinal (LO) phonon modes, and to the fine structure of spectra in the frequency region between them ({\it Restrahlen} band).   Reviews of earlier experimental and theoretical works in this area can be found in Refs.\onlinecite {Sievers,Taylor-review}. In spite of the disordered nature of mixed crystals, it is usually possible to observe both TO and LO modes of pure crystals at both ends of the concentration range, as well as features associated with disorder.\cite{Sievers,Taylor-review,Chang-Mitra} This is usually done in reflectance or transmittance experiments by identifying  the maxima of $ Im(\epsilon)$ and $ -Im(1/\epsilon)$ with TO and LO modes respectively, where $\epsilon$ is the dielectric function.\cite{Moss}

The term mixed crystals is usually applied to materials in which the concentration of each component is large, so that no component can be considered as a system of independent impurities. At  smaller concentrations the dynamics of impure systems is described in terms of defect states, which are either localized or quasilocalized depending on whether they fall into forbidden or allowed bands\cite{Lifshitz,Lifshitz-Kosevich,Maradudin,Marad-Montroll} of the pure system. The transition from low to high concentration behavior occurs when the average distance between defects becomes of the same order as the localization length of the individual states. In the case of the local phonon states, the localization length is of the order of a few lattice constants, and the respective transition concentration is of the order of $\sim 10^{18}$ $cm^{-3}$. Most of the studies of mixed crystals were focused upon combinations of alkali halides, II/IV and III/V group polar binary crystals $AB_{1-p}C_{p}$, where $p$ is concentration. The $B$ atoms are substituted with the atoms $C$ from the same column of the Periodic Table.  It was found that there were several patterns that TO and LO modes follow in mixed crystals.\cite{Sievers,Taylor-review,Chang-Mitra,Lucovsky,Harada-Narita} In the type I mixed crystals, also called one-mode crystals, the frequencies of the TO and LO modes evolve smoothly and almost linearly with the composition parameter $p$ from their values in $AB$ to those in $AC$. In some one-mode systems (e.g., $Ba/SrF_{2}$, $Ca/SrF_{2}$\cite {BaSrF2Verleur,BaSrF2Chang}) an additional pair of weak modes inside the absorption band is observed. 

Type II or two-mode mixed crystals exhibit qualitatively different behavior. Two distinct restrahlen bands are usually observed in the crystals of this category. For arbitrary concentration, the width of each of these  bands, $\Omega_{LO}-\Omega_{TO}$, is approximately proportional to the composition parameter $p$ for the dopant related band, and to $1-p$ for that of the original material.  Two-mode behavior is characteristic for the crystals formed by elements from III and V groups of the Periodic Table (e.g. $GaAs/P$ and the like).

Another mode behavior could not be fit experimentally either to one- or to two-mode patterns (e.g. $Ga/InP$,\cite {GaInP}, $PbSe/Te$,\cite{PbSeTe} $K/RbI$,\cite {KRbIFertel,KRbIRenker} {\it etc.}). This group of mixed crystals was called Type I-II or one-two-mode systems.  In these systems, one mode behaves as in Type I, while the other behaves as in Type II, which is why Type I-II is also called mixed mode. 

To explain different types of behavior in mixed polar crystals, a number of theoretical models have been proposed. In the IR experiments, modes with wave numbers $q\sim 0$ are excited. The Random Element Isodispacement model (REI)\cite{Chen,Verleur} takes advantage of this fact, assuming that the sublattices vibrate in phase with $q=0$. With some modifications\cite{Chang-Mitra} after including the local field, and assuming a dependence of the force constants upon the composition, the modified REI model (MREI) was successfully used to fit two-mode crystals but encountered some problems in the one-mode materials. The fine structure of some mixed crystals was reproduced along the lines of REI by means of a considerable increase in the number of fitting parameters (cluster model\cite{Verleur}). A different approach made use of the Green's function formalism in order to treat vibrations in the mixed crystals; averaging over disorder was performed either in a simple virtual crystal approximation (VCA) or with the use of the much more elaborate coherent potential approximation\cite{Elliott} (CPA). Interaction with the electromagnetic field in these approaches was included after the averaging procedure was performed. The mean field approximation treats the crystal as if it has atoms of the same sort with some average properties. It reproduces well the one-mode behavior, but could not possibly explain  the two-mode systems. CPA combined with an electrostatic treatment of the electric field was used to fit a broader range of experimental data (see, for instance, Ref. \onlinecite {Bonneville}).

An important theoretical problem many researchers have focused upon is to find a simple criterion for different mode behavior, based upon dynamical properties of the crystals.  Lucovsky\cite{Lucovsky} originally considered the significance of whether the original absorption bands overlap or not, which turned out to be a rather rough criterion. Harada and Narita\cite{Harada-Narita} proposed a criterion based on relations between MREI constants (see also review papers in Refs.\onlinecite{Sievers,Taylor-review}). It turned out that one-mode and two-mode crystals can be described using quite simple models with only a few fitting parameter, such as VCA and MREI, respectively. Lucovsky {\it et al.} in Ref. \onlinecite {Lucovsky} made an attempt to establish a connection between the behavior of a mixed crystal and the existence of the localized modes at small values of concentration parameters $p$ or $1-p$. It was suggested that the two-mode behavior is associated with the existence of local impurity states at both ends of the concentration range, while the one-mode regime occurs when such local states do not arise.  These ideas, however, could not be applied to the mixed-mode behavior since the local phonon states arise either in the gap between acoustic and optic branches or above the optic branch of the phonon spectrum.\cite {Lifshitz,Lifshitz-Kosevich,Maradudin,Marad-Montroll} Therefore, the mixed-mode crystals required much more elaborate models, where agreement with experimental data can only be achieved by increasing the number of fitting parameters. The same was true for weak features of spectra of the one-mode crystals.

One of the goals of the present paper is to put forward a simple physical picture explaining the mixed-mode behavior based upon the concept of local polaritons and an impurity-induced polariton band. The local polariton is a state that arises due to an impurity with frequencies inside the {\it restrahlen} band of a crystal. These states have an electromagnetic component which is localized around the defect, because it cannot propagate in the region with a negative dielectric parameter. A local polariton state within the {\it restrahlen} region was first observed experimentally in Ref. \onlinecite{Dean} as a weak feature in reflectance and Raman spectra of $GaAs$ doped with $Sn,Te$ or $S$. These states were found to be associated with the local change in susceptibility due to localization of an electron around the dopant. An interaction between the localized electron and LO vibrations of the crystal gives rise to local LO phonons, therefore these local states have a complicated structure that involves interactions between electrons, phonons and electromagnetic field. Theoretically these states, however, were only considered in the electrostatic approximation,\cite{Dean} and therefore these early observations did not lead to the concept of local polaritons, which requires that the retardation be taken into account. 

A new wave of interest in the optical properties of impure crystals in the {\it restrahlen} region occured recently and was due to a general interest in systems with depleted or altered electromagnetic density of states (DOS). The primary examples of such systems are photonic crystals,\cite{Photonics-Book} and microcavities.\cite{microcavities} The {\it restrahlen} region in polar crystals was considered from this new perspective independently in Refs.\onlinecite{Rupasov,Classics,Deych}, where the concept of local polaritons was introduced. The local state considered in Ref.\onlinecite{Rupasov} arises due to an impurity atom with optically active electronic transitions, which interacts with host atoms through the electromagnetic field only. With the transition frequency inside the {\it restrahlen} band, this atom forms a local atom-radiation bound state. To some extent these states are similar to the ones observed by Dean {\it et al.},\cite{Dean} though an interaction with phonons was left out in  Ref.\onlinecite{Rupasov}. A different type of local polariton was considered in Refs.\onlinecite{Classics,Deych}, where it was shown that regular isotopic impurities can give rise to local states with frequencies inside the {\it restrahlen} region. The interaction with the retarded electromagnetic field is responsible for the localized electromagnetic component of the states. The detailed analysis of the local polariton mode in a three dimensional sodium-chloride-like structure (BCC) crystal in the case of diagonal and off-diagonal disorder was carried out in Ref. \onlinecite {Podolsky}. It was proven that, because of retardation effects, a local polariton mode  splits off the bottom of the TO-LO gap for an arbitrarily small strength of the defects even in three dimensional systems.


When considering local polaritons, the authors of Refs. \onlinecite {Classics,Deych,Podolsky} assumed that the {\it restrahlen} is  a spectral gap not only for electromagnetic excitations but for phonons as well. This is indeed the case for some polar crystals. It occurs more frequently, however, that the {\it restrahlen} region is devoid of transverse optic phonons, but is still filled with LO modes. In this case the interaction between local polaritons and LO phonons would make the former quasi-stationary. Whether the local polaritons survive this interaction depends upon their life-time, and we shall address this question  in the present paper. At the same time it is useful to note that the states observed in Ref. \onlinecite{Dean} reside in the frequency region where the density of LO modes is especially large. The fact that they remained observable allows for optimism that the local polariton states of Refs.\onlinecite{Classics,Deych,Podolsky} could also survive provided that DOS of LO is not too large. If this is the case, then local polaritons due to isotopic impurities can be invoked to explain the mixed-mode behavior of mixed crystals and weak features in the one-mode systems without having to introduce tens of fitting parameters. 

These ideas can, for example, be used to discuss the one-two-mode mixed crystal $Ga_{0.70}In_{0.30}P$. Pure $GaP$ has a complete polariton gap in all directions, and experiments of Ref. \onlinecite{Livescu} clearly demonstrated the polariton band associated with $In$ atoms inside the {\it restrahl} of $GaP$ even at room temperature. Polariton branches inside {\it restrahlen} were also observed in pure $CuCl$.\cite{Livescu} This material demonstrates a very peculiar behavior because $Cu$ atoms can occupy several non-equivalent positions, thereby producing defects even in a pure material.\cite {Lucovsky,OffCenter} These defects were shown to be responsible for a new TO mode with a frequency in the main {\it restrahlen}.\cite {Lucovsky,OffCenter} These experiments were originally described using a phenomenological two-oscillator model, but they perfectly fit to the idea of local polariton sates. 

Local polaritons, however, can  take us further than an explanation of the old reflectivity experiments.  When retardation is taken into account, the local polaritons give rise to a new transmitting channel for electromagnetic excitations, via  impurity-induced polariton band.
Experimentally, such a band can be observed in transmission steady-wave and time-resolved experiments. The majority of the old experiments mentioned above dealt with reflection spectra. Dips in the reflection coefficient in those experiments were associated with impurity-induced absorption, while they could have actually occured because of enhanced transmission via the impurity-induced polariton band. The effect of the enhanced transmission inside the {\it restrahlen} region of a polar crystal was observed in $CuCl$.\cite{CuClTransm}

We showed analytically that local polaritons give rise to a strong resonance enhancement of the transmittance at the frequency of the local mode,\cite{EuroPhysLett,One-Imp-Transm} using a one-dimensional chain of dipoles. In the system without dissipation, the maximum value of the transmission coefficient reaches unity, although peak width depends exponentially upon the size of the system. Numerical simulations for a finite concentration of impurities showed that a localized mode develops into a conduction band inside the original polariton gap.\cite{One-Imp-Transm} The growth of the local  state into a conduction band in the system of optically active defects was also demonstrated in Ref. \onlinecite{Singh}.
An analytical treatment of the impurity-induced polariton band in the dispersionless one dimensional dipole chain, with concentration of defects varying in the range $0<p<1$ was proposed in Ref. \onlinecite{Lyapunov}. Using the microcanonical method\cite{MicrocanMethod} we were able to calculate DOS due to the defect subsystem and the extinction (localization) length. We found that the impurity-induced band has a number of interesting properties. The group velocity of electromagnetic excitations, propagating in this band, for instance, was found to be proportional to the concentration of the impurities, and can be significantly smaller than the speed of light in vacuum.  The results of analytical calculations were in agreement with numerical simulation in the presence of the spatial dispersion. The knowledge obtained from analytical dependencies of DOS in the one-dimensional model gives us insight into a treatment of three dimensional  mixed polar crystals. The spatial size of local polariton states may be as large as the EM wavelength, meaning that even at a very low impurity concentration, $\sim 10^{12}$ $cm^{-3}$, the local polaritons significantly overlap. This fact allows us  to develop a continuous  approximation for calculating properties of the polariton-induced band. Applying this approximation to the one-dimensional case, we can use our previous one-dimensional results to test the suggested approach.

The structure of our paper is as follows. In Section II, the continuous approximation for the mixed crystals, based upon the concept of local polaritons, is introduced. In Section III, we estimate the life-time of the polariton local states due to interaction with LO modes. We show that under realistic assumptions about the density of states of LO modes, this life-time is compatible with that caused by inelastic relaxation, and may be long enough for local polaritons to survive. Section IV deals with the properties of the impurity-induced polariton band (dispersion laws and density of states of the respective excitations are studied). We also consider in this section the concentration dependence of poles and zeroes of the dielectric function, and a simple explanation of the mixed-mode and one-mode (with weak structure) behaviors is proposed. Section V is devoted to investigation of the reflection and transmission spectra of the semi-infinite mixed crystals and a slab of finite width. In Section VI we calculate  scattering length of the impurity induced polaritons due to the fluctuations of concentration, and demonstrate why this scattering is significantly less important for impurity-induced polaritons then for regular polariton branches. This section provides an additional justification for the main results of the paper.  

\section{Model}

In our recent work, Ref.\onlinecite{Lyapunov}, we considered development of an impurity-induced polariton band in a one-dimensional model, and used the so-called microcanonical method\cite{MicrocanMethod} in order to analytically calculate the complex Lyapunov exponent of the system. The approximation used turned out to be in a very good agreement with numerical results, and was shown to be equivalent to a continuous medium approximation.  The latter can  easily be generalized for 3-d systems, and below we develop an approach to the impurity-induced polariton band in 3D mixed polar crystal using the following fundamental assumptions. 

In order to model optic vibrations of the system under study, we introduce two subsystems of oscillators with different frequencies. One subsystem represents an optic mode of the host atoms.  The interaction between this mode and light leads to the polariton gap between TO and LO frequencies of the pure crystal. The second subsystem introduces vibrations of impurities, and it is assumed that its frequency belongs to the {\it restrahlen} of the host crystal. This is a crucial assumption of the model, since interaction between light and this mode brings about local polariton modes. The significance of this assumption rests upon two ideas. The first one goes back to Lucovsky,\cite{Lucovsky} who was the first to connect modes of mixed crystals with local impurity modes. The second one is the concept of local polaritons, which are local states with frequencies in the restrahlen of the host. In the one-dimensional case we have confirmed analytically and numerically\cite{One-Imp-Transm,Lyapunov} that the presence of the local polaritons gives rise to an impurity polariton branch, when the concentration of impurities grows.

 The frequencies of both oscillators are, in general, complex-valued. The imaginary part of host vibrations is due to anharmonicity, while the impurity mode has two sources of decay. Besides usual relaxation, local polaritons can acquire an additional imaginary part due to interaction with LO phonons which in some cases can fill the {\it restrahlen}. We will consider the contribution of this interaction into the life-time of local polaritons in the next section, and the general effects of relaxation upon reflection spectra of mixed crystals will be discussed in Section V.

It was found in Refs.\onlinecite{Classics,Deych,Podolsky,Lyapunov} that for all frequencies outside of the immediate vicinity of the TO boundary of the {\it restrahlen}, the localization length $l_{0}$ of local polaritons is of the order of magnitude of the wavelength of the incident light $\sim 10^{-3}\ cm$. Even for residual  concentration $n \sim 10^{12}\ cm^{-3}$ of the impurities, $nl_{0}^{3}$ is  large. The individual states significantly overlap and  a macroscopic volume containing many impurities can still be much smaller than the localization length. This fact allows us to develop a continuous medium approximation similar to the usual one used to treat long-wave excitations,\cite{Born-Huang} but this time we use it to treat the sub-system of impurities. We would like to emphasize again that this consideration is reasonable for the polariton band only because of the long-range nature of the localized electromagnetic component of the local polaritons.  

The microscopical Hamiltonian describing optic modes of our system is:
\begin{equation}
H=\sum_{{\bf r}}\frac{\mu _{0}}{2}\left( \frac{d{\bf u}_{\bf r}}{dt}\right) ^{2}+
\sum_{{\bf r'}}\frac{\mu _{1}}{2}\left( \frac{d{\bf v}_{\bf r'}}{dt}\right)^{2}+
\frac{1}{2}\sum_{{{\bf r},{\bf r'}}} \Phi \left( {\bf u}_{\bf r}, {\bf v}_{\bf r'} \right) ,
\label{H_initial}
\end{equation}
where $\mu_{0}, \ \mu_{1}$ and ${\bf u}_{\bf r}, \  {\bf v}_{\bf r'}$ are the reduced masses and relative displacements of the ions in pure $AB$ and $AC$ crystals, respectively. We will assume that the presence of atoms $C$ does not affect the order in the $A$ sublattice. The summation runs over spatial index ${\bf r}$ for host atoms and over ${\bf r'}$ for the impurities. The third term in the above expression is the potential energy. In the  harmonic approximation it can be written as 
\begin{eqnarray}
\Phi \left( {\bf u_r}, {\bf v_{r'}} \right)&=&
\Phi_{i,j}^{(0)} \left( {\bf r} - {\bf r}_1 \right) u_{\bf r}^{i} u_{{\bf r}_1}^j+
\Phi_{i,j}^{(1)} \left( {\bf r'} - {\bf r}_1' \right) v_{\bf r'}^i v_{{\bf r}_1'}^j
\nonumber \\
&+&\Phi_{i,j}^{(01)} \left( {\bf r} - {\bf r'} \right) u_{\bf r}^i v_{\bf r'}^j.
\label{Phi}
\end{eqnarray}
In order to avoid unnecessary complications, we assume no spatial dispersion  in our model Hamiltonian (\ref{H_initial}) and isotropy of force constants
\begin{equation}
\Phi \left( {\bf u}_{\bf r}, {\bf v}_{\bf r'} \right)=\Phi_{0}{\bf u}_{\bf r}^{2}+\Phi_{1}{\bf v}_{\bf r'}^{2} ,
\label{PhiDispLess}
\end{equation}
where $\Phi_{0}$ and $\Phi_{1}$ are force constants describing the interaction of the ions in the $AB$ and $AC$ lattices, respectively. As follows from the form of expression (\ref{PhiDispLess}), the contributions due to host and impurity ions in the Hamiltonian (\ref{H_initial}) can be separated
\begin{eqnarray}
H&=&\sum_{{\bf r}}\left[ \frac{1}{2}\mu_{0}\left( \frac{d{\bf u}_{\bf r}}{dt}\right)^{2}+\frac{1}{2}\Phi_{0}{\bf u}_{\bf r}^{2} \right] 
\nonumber \\
&+&\sum_{{\bf r'}}\left[\frac{1}{2}\mu_{1}\left( \frac{d{\bf v}_{\bf r'}}{dt}\right)^{2}+\frac{1}{2}\Phi_{1}{\bf v}_{\bf r'}^{2} \right].
\label{H_separated}
\end{eqnarray}
We have several  parameters of the dimension of length in our system: a lattice constant, $a$; an average distance between defects, $l \sim n^{-1/3}$, which depends on concentration; and the localization length and the wavelength of the incident light, which are of the same order, $l_{0}\sim \lambda \sim 10^{-5} \ m$. We shall assume that the concentration is such that $n \gg l_{0}^{-3}$. When this condition is satisfied, the individual defect states overlap, and the dynamical properties, apart from those at the fluctuation band-edge, do not depend significantly upon the particular arrangement of the defects in the crystal. Hence, one can introduce a smoothing parameter $l_{\delta V}$ such that $l \ll l_{\delta V} \ll l_{0}$. The macroscopic volume associated with this parameter is such that it contains macroscopically many impurities, but there are still no significant changes in spatial distribution of vibrations over this length: $ {\bf u}_{\bf r+R}, {\bf v}_{\bf r+R}\approx {\bf u}_{\bf r}, {\bf v}_{\bf r}$ for $R \sim l_{\delta V}$. Taking advantage of this fact one can sum over the volume $\delta V({\bf R}) \sim l_{\delta V}^{3}$ in the vicinity of ${\bf R}$ in the Hamiltonian (\ref{H_separated}): 
\begin{eqnarray}
\sum_{ {\bf r} \in \delta V({\bf R}) }
\left[ \frac{1}{2}\mu_0\left( \frac{d{\bf u_r}}{dt}\right)^2+\frac{1}{2}\Phi_0{\bf u_r}^2 \right] &\rightarrow &
\nonumber \\
\rightarrow \left[ \frac{1}{2}\mu_{0}\left( \frac{d{\bf u_R}}{dt}\right)^2+\frac{1}{2}\Phi_0{\bf u_R}^2 \right] 
\cdot &n_0& \left[1 - p({\bf R}) \right] \delta V ({\bf R}),
\label{Smoothing1}
\end{eqnarray}
\begin{eqnarray}
\sum_{{\bf r'} \in \delta V({\bf R})}\left[ \frac{1}{2}\mu_0\left( \frac{d{\bf u_{r'}}}{dt}\right)^{2}+\frac{1}{2}\Phi_0{\bf u_{r'}}^{2} \right] &\rightarrow &
\nonumber \\
\rightarrow \left[ \frac{1}{2}\mu_0\left( \frac{d{\bf u_R}}{dt}\right)^2+\frac{1}{2}\Phi_0{\bf u_R}^{2} \right] &\cdot & n_0 p({\bf R})  \delta V ({\bf R}),
\label{Smoothing2}
\end{eqnarray}
where we assume that the change in lattice constants of two end crystals is negligible, $a \simeq n_{0}^{1/3}$, and $p({\bf R})=n({\bf R})/n_{0}$. Thus the Hamiltonian (\ref{H_separated}) can be rewritten in the continuous medium approximation as
\begin{eqnarray}
H= \frac{n_{0}}{2} \int \left[ \left(1-p({\bf R})\right) \mu_{0} \stackrel{\bf .}{\bf u}_{\bf R}^{2} + (1-p({\bf R})) \Phi_{0}{\bf u}_{\bf R}^{2} \right.
\nonumber \\
\left. +p({\bf R})\mu _{1}\stackrel {\bf .}{\bf v}_{\bf R}^{2}+p({\bf R})\Phi_{1}{\bf v}_{\bf R}^{2}\right]dV.
\label{H_longwave}
\end{eqnarray}
The ion polarization at ${\bf R}$ is
\begin{eqnarray}
{\bf P}_{ion}({\bf R}) &=& \frac{1}{ \delta V({\bf R}) } \left[ \sum_{ {\bf r} \in \delta V({\bf R}) } q {\bf u}_{\bf r} + 
                                                       \sum_{ {\bf r'} \in \delta V({\bf R}) } q {\bf v}_{\bf r'} \right] 
\nonumber \\
&\rightarrow &
(1-p({\bf R})) q n_{0} {\bf u}_{\bf R}+ p({\bf R}) q n_{0} {\bf v}_{\bf R}\ ,
\label{P_ion}
\end{eqnarray}
where we assume for simplicity the same effective charges for both oscillators, and again use the fact that $l_{\delta V} \ll l_{0}$ and consequently ${\bf u}_{\bf r},\ {\bf v}_{\bf r'}$ do not change significantly over such distances. The interaction of the polarization with the electromagnetic field gives rise to an additional term in the Hamiltonian:
\begin{equation}
U_{int}= \int {\bf P}\cdot {\bf E}dV 
           = n_{0} \int\left[ (1-p({\bf R})) {\bf u}_{\bf R}+ p({\bf R})  {\bf v}_{\bf R} \right] \cdot {\bf E}dV.
\label{U_int}
\end{equation}
Combining Eqs. (\ref{H_longwave}) and (\ref {U_int}) and writing out the resulting Hamilton equations for ${\bf u}_{\bf R},\ {\bf v}_{\bf R}$, one obtains
\begin{eqnarray}
  \mu_{0} \stackrel{..}{\bf u}_{\bf R}=\Phi_{0} {\bf u}_{\bf R}+q{\bf E}, 
\nonumber 
\\
  \mu_{1} \stackrel{..}{\bf v}_{\bf R}=\Phi_{1} {\bf v}_{\bf R}+q{\bf E}. 
\label{Eq_motion}
\end{eqnarray}
These equations of motion should be accompanied by the Maxwell equation for the electromagnetic field in the medium
\begin{equation}
{\bf \nabla} \times{\bf \nabla} \times {\bf E} = \frac{1}{c^{2}} \frac{d^{2} \left( {\bf E} +4 \pi {\bf P}_{ion}+4 \pi {\bf P}_{el} \right) }{dt^{2}},
\label{Maxwell}
\end{equation}
where the last term describes an electron contribution to the polarization that determines the high-frequency value of the dielectric parameter. Another effect to be taken into account is that the local electric field on atoms that induces polarization is different from the macroscopic field entering the Maxwell equations. The effective local field in a high symmetry crystal can be written in the form:\cite{Born-Huang,Bonneville}
\begin{eqnarray}
{\bf E}_{loc} & =& {\bf E} + \frac{4 \pi}{3} \left( {\bf P}_{ion} + {\bf P}_{el} \right) ,
\label{E_loc} \\
{\bf P}_{el} &=& n_{0} {\alpha}_{\infty} {\bf E}_{loc}.
\label{P_el}
\end{eqnarray}
Substituting the effective local field into Eq. (\ref{Eq_motion}) and total polarization of the volume ${\bf P}={\bf P}_{ion}+{\bf P}_{el}$ into Eq. (\ref{Maxwell}), one obtains the closed system of equations of mechanical coordinates, the polarization and the electric field. Eqs. (\ref{Eq_motion}), (\ref{Maxwell}), (\ref{E_loc}), and (\ref{P_el}) formally resemble equations of MREI used in many papers. There are, however,  important differences. First, the concentration parameter $p({\bf r})$ entering our equations is still a random function of coordinates. The statistics of this parameter, however, are significantly different from the original statistics of the microscopic distribution of impurities. The spatial inhomogeneity of the continuous function $p({\bf r})$ reflects the fluctuations in the number of impurities in the macroscopic volume $\delta V$, which are significantly reduced as compared to the fluctuations in the original microscopic distribution of impure atoms. More detailed consideration of the statistical properties of $p({\bf r})$ will be presented in Section VI. Second, the derivation of Eqs. (\ref{Eq_motion}), (\ref{Maxwell}), (\ref{E_loc}), and (\ref{P_el}) is explicitly based upon existence of a macroscopic length scale -- the localization length of the local polaritons. A similar procedure cannot be applied to the regular phonon states, because the localization length of the local phonon states is usually microscopic. Finally, the requirements that the impurity-induced oscillator has its frequency within {\it restrahlen} replaces conditions for local phonon modes that appear in regular MREI.\cite{Chang-Mitra}
In this section we will assume that the fluctuation of the number of defects in the chosen volume is negligible and, therefore, $p({\bf R}) \equiv p$. Excluding the polarization and lattice displacements from the system yields the following equation for the Fourier components of the electric field 
\begin{equation}
{\bf k} \times{\bf k} \times {\bf E}_{\omega} = \frac{{\omega}^{2}}{c^{2}} \epsilon (\omega) {\bf E}_{\omega} .
\label{Maxwell_final}
\end{equation}
In this equation, $\epsilon (\omega)$ denotes the effective dielectric function of the mixed polar crystal $AB_{1-p}C_{p}$
\begin{equation}
\epsilon (\omega) = {\epsilon}_{\infty}  \frac
{1+{\displaystyle \frac{2}{3}} \left( (1-p){\displaystyle \frac{d_{0}^{2}} {{\Omega}_{0}^{2}-{\omega}^{2}}}+p{\displaystyle \frac{d_{1}^{2}}{{\Omega}_{1}^{2}-{\omega}^{2}}} \right)}
{1-{\displaystyle \frac{{\epsilon}_{\infty}}{3}} \left( (1-p){\displaystyle \frac{d_{0}^{2}}{{\Omega}_{0}^{2}-{\omega}^{2}}}+p{\displaystyle \frac{d_{1}^{2}}{{\Omega}_{1}^{2}-{\omega}^{2}}} \right)},
\label{Epsilon}
\end{equation}
where $\Omega_{0,1}^2= ( \Phi_{0,1}/\mu_{0,1} )^{1/2}$ are the lattice eigenfrequencies and parameters 
\[
d_{0,1}^{2}=4 \pi {\displaystyle \frac{{\epsilon}_{\infty}+2} {3{\epsilon}_{\infty}}}{\displaystyle \frac{q^{2}}{\mu_{0,1}}}
\]
determine the width of the polariton gaps of the end crystals. We assume that the high frequency dielectric constant does not depend upon the concentration. $\epsilon_{\infty}$ can be expressed in terms of polarizability $\alpha_{\infty}$ as 
\[
\epsilon_{\infty} =1+{\displaystyle \frac{4 \pi \alpha_{\infty} n_{0}}{1-\frac{4\pi}{3} \alpha_{\infty} n_{0}}}.
\] 

From Eq. (\ref{Maxwell_final}) one can separate the transverse  and  longitudinal modes as
\begin{eqnarray}
k^{2} {\bf E}_{\omega}^{\perp} &=& \frac{{\omega}^{2}}{c^{2}} \epsilon (\omega) {\bf E}_{\omega}^{\perp}\ ,
\label{Maxwell_perp} \\
\epsilon (\omega) {\bf E}_{\omega}^{\parallel}&=&0 \ .
\label{Maxwell_parallel}
\end{eqnarray}
A similar expression for the dielectric function was obtained previously in Ref.\onlinecite{Harada-Narita} within the MREI. Nobody, however, assumed before that the $\Omega_1$ lies within the {\it restrahlen} of the host. It is only the concept of local polaritons that justifies using this dielectric function  in the case when the frequency of the defect oscillator falls into the polariton band gap. Comparing Eq. (\ref{Epsilon}) with the respective expression of our earlier work on the 1-d model Ref.\onlinecite{Lyapunov}, one can see that Eq. (\ref{Epsilon}) is a zero-order approximation in a series expansion in terms of the  small parameter $l/l_{0} \ll 1$. In Section VI of the paper we shall demonstrate this fact implicitly.

\section {Long-living quasistationary local polariton states in restrahlen band}

The model developed in the previous section is implicitly based upon the assumption about the existence of local states with frequencies within the {\it Restrahlen}. In this section we shall provide additional justifications for this assumption. 

In the original papers\cite{Classics,Deych} where local polaritons were introduced, it was assumed that within the restrahlen there is a genuine spectral gap with the phonon DOS being equal to zero over some frequency region. Such situations are, indeed, possible in some crystals ($GaP$, $ZnS$, $CuBr$, {\it etc}.\cite {Phonon-dispersion}). It is more common, however, that the {\it restrahlen} is filled with LO phonons, which are linearly coupled to local polaritons. This coupling results in the ``phonon radiative decay" of the local polaritons (compare to radiative decay of local phonons due to coupling to light). It is important to emphasize, however, that the electromagnetic component of the local polaritons remains localized even when there is leakage through the phonon component. The phonon radiative decay broadening of local polaritons (and the respective life-time) is determined by the density of LO phonon states within the {\it restrahlen}. In some cases this DOS is large enough to suppress the local polaritons.  At the same time, there exists a broad range of materials (for instance, $NaF$,$NaBr$, $RbF$, {\it etc}.\cite {Phonon-dispersion}) that have a relatively low DOS inside the {\it restrahlen}, so one could expect that local polaritons can survive the presence of LO phonons and affect significantly  the transport properties of the system. In fact, the local states inside the {\it restrahlen} observed in Ref.\onlinecite{Dean} resided in the frequency region with a rather large DOS, and still could be observed both in reflection and Raman spectra.

Since the fundamental assumptions incorporated into the model presented in this paper are based upon existence of long-living local polaritons, it is of great importance to study their life-time in realistic polar crystals. Our consideration of the question about the ``phonon radiative decay" time of the local polaritons is based upon results of Ref.\onlinecite {Podolsky}. It was shown that the frequency of the local polariton is determined by
\begin{eqnarray}
1 &=& - \frac {\delta m}{m}\omega ^{2}\int \frac {\rho _{\parallel }(z)+2\rho _{\perp }(z)}{\omega ^{2}-z^{2}}dz
\nonumber
\\
&=& -\frac {\delta m}{m}\left[ \omega ^{2}\int \frac {\rho _{\parallel }(z)+2\rho _{\perp }(z)}{\omega ^{2}-z^{2}}dz\ \right] ,
\label{Eigenmodes_d}
\end{eqnarray}
where $m=\frac{m_{-}+m_{+}}{m_{-}/m_{+}}$, $m_+$ and $m_-$ are the masses of positive and negative ion sublattices respectively, $\delta m$ is the difference between masses of the defect and the host atom replaced by the defect, and $\rho_{\parallel}(\omega)$ and $\rho_{\perp}(\omega)$ are densities of states of TO and LO phonons, respectively. If the LO DOS differs from zero in the {\it restrahlen} region where we expect the local state to reside, the respective integral in Eq.(\ref{Eigenmodes_d}) acquires an imaginary part equal to $\ i\pi \omega \rho _{\parallel }(\omega )$. For a light impurity ($\delta m<0$) this equation  would have a real solution $\omega _{l}$ if $\rho_{\parallel}(\omega)=0$.{Podolsky} The presence of LO modes makes the solution complex valued,  $\omega' _{l}=\omega_l - i\gamma$. The phonon component becomes delocalized, although the electromagnetic component remains localized. 

If, as we assume, $\gamma /\omega _1\ll 1$, then $\omega _1$ can be considered to be the solution of Eq. (\ref{Eigenmodes_d}) with the imaginary part dropped
and $\gamma $ can be found as the first order correction to it 
\begin{equation}
\frac {\gamma }{\omega _{1}}=\frac {\pi \omega _{1}\rho _{\parallel }(\omega _{1})}{2\omega _{1}^{2}R(\omega _{1})+\omega _{1}^{3}\frac {dR(\omega _{1})}{d\omega_1 }},
\label{gamma_1}
\end{equation}
where $R(\omega_1)$ is a principal value of the integral entering Eq.(\ref{Eigenmodes_d}). The derivative of $R(\omega_1 )$ can be estimated using the explicit form of this function as $\omega _{1}\frac {dR(\omega _{1})} {d\omega_1 }=2R(\omega _{1})+O\left((\omega _{1}a/c)^2 \right)$. Finally one obtains
\begin{equation}
\frac {\gamma }{\omega _{1}}=\frac {\pi }{4}\left( \frac {-\delta m}{m_{-}+m_{+}}\right) \frac {m_{-}}{m_{+}}\omega _{1}\rho _{\parallel }(\omega _{1}).
\label{gamma}
\end{equation}
It follows from Eq. (\ref {gamma})  that there are three factors affecting the life-time of the polariton states.
First of all, it is the density of LO phonon states $\omega _{1}\rho _{\parallel }(\omega _{1})$ in the {\it restrahlen}. As  mentioned above, due to the strong dispersion of the LO branch, the density of states in many alkali halides\cite {Phonon-dispersion} between $\Omega _{(TO)}(0)$ and $\Omega _{(LO)}(0)$ can be less then $10\%$ of the maximum DOS at TO frequency. Unfortunately, the experimental data on the phonon DOS known to us do not provide its value  in absolute units, and in order to obtain an estimate for this quantity in the region of interest, we have to rely upon some assumptions. To this end we use the Debye model for DOS of the {\it acoustic} phonons in order to establish a scale for the experimental results listed in Ref. \onlinecite{Phonon-dispersion}. This choice seems reasonable because the Debye model gives a fairly good description of low-temperature thermodynamic properties of crystals, and parameters of the model for most of the crystal of interest are established with good accuracy. Having the scale for the phonon DOS, one can assess the numerical value of the dimensionless quantity $\omega _{1}\rho _{\parallel }(\omega _{1})$. Our estimates show that this parameter in the spectral region of interest can take values between $0.2$ and $0.6$ in different systems.

The next factor affecting the life-time of the local polaritons is the defect parameter $\delta m$. This parameter cannot be assumed to be too small since we would like to have our local state farther away from the TO boundary of the {\it restrahlen}, and as it was shown in Refs. \onlinecite {Classics,Podolsky}, the frequency of the  local polariton state moves deeper into the {\it restrahlen} with increase of the defect strength $\delta m$. At the same time, it is clear that the factor $-\delta m/(m_{-}+m_{+})$ is at least less then unity. 
The third factor $m_{-}/m{+}$ (or $m_{+}/m{-}$ when the negative atom is replaced), can be as small as $0.2$ for $RbF$ or even $0.06$ in the case of $LiI$. Combining all terms we find that in a number of crystals ($NaBr$, $NaCl$, $RbF$, {\it etc},) the dissipation of the local polariton state is rather small $\gamma/\omega _1 < 0.1$. In fact, it appears to be of the same order of magnitude as anharmonic absorption, and later in the paper we will  introduce both these relaxation channels phenomenologically using just one parameter of relaxation. The presented estimates show that local polaritons can actually survive even in materials with {\it restrahlen} filled with LO phonons, and provide, therefore, a foundation for the model presented in the previous section. Now we can start discussing the results following from this model.

\section {Impurity-induced polariton band in  mixed polar crystals}

\subsection {Dispersion laws of impurity polaritons}

In this section we discuss properties of polariton excitations in our system neglecting relaxations. Effects of the dissipation will be incorporated in our treatment of reflection spectra in the next section. The dispersion of the transverse polaritons is determined by the equation
\begin{equation}
k=\sqrt{\epsilon (\omega)}\frac{\omega}{c},
\label{transv_k}
\end{equation}
while the longitudinal excitations obey the equation $\epsilon (\omega)=0$ and are dispersionless within the present model. The sign of $\epsilon (\omega)$ determines  the structure of the spectrum. The bands of propagating electromagnetic waves coupled to the lattice vibration appear at frequencies where $\epsilon (\omega)$ is positive, and bandgaps arise where the dielectric function (\ref{Epsilon}) is negative. The change of the sign of $\epsilon (\omega)$ occurs when $\epsilon( \omega ) = 0$ and when $1/\epsilon( \omega ) = 0$. In the electrostatic approximation, the first of these conditions determines the LO frequencies $\omega_{LO}$, while the second gives the frequencies of TO phonon modes, $\omega_{TO}$. With retardation taken into account for regular polaritons in pure crystals, the latter becomes  the short wave limit of the lower transverse  polariton branch, while $\omega_{LO}$ determines the $k=0$ frequency of the upper polariton transversal branch degenerate with the longitudinal phonon mode. As we shall see in this section for the impurity induced polaritons, the interpretation of the boundary frequencies is quite different.

In the absence of defects, $p=0$, the {\it restrahlen} polariton band stretches between transverse , 
\[
\Omega_{0}^{(TO)2}=\Omega_0^2-\displaystyle{\frac{\epsilon_{\infty}}{3}d_0^2},
\]
 and longitudinal, 
\[
\Omega_{0}^{(LO)2}=\Omega_{0}^{2}+\frac{2}{3}d_{0}^{2},
\]
 frequencies. Introducing defects with $\Omega_{0}^{(TO)2} < \Omega_{1}^{2} < \Omega_{0}^{(LO)2}$ in the system with concentrations significant enough to satisfy the condition $l/l_{0} \ll 1$  yet still small in the sense that $ p \ll 1$, one can rewrite Eq. (\ref{Epsilon}) in the linear in $p$ approximation:
\begin{equation}
\epsilon (\omega) \simeq \epsilon_{\infty} \left( \frac{\Omega_{0}^{(LO)2}-\omega^{2}} {\Omega_{0}^{(TO)2}-\omega^{2}}\right)
\frac {\omega _{il}^{2}- \omega ^{2} } {\omega _{iu}^{2} - \omega ^{2} }.
\label{Epsilon_small_p}
\end{equation}
It can be directly seen from this expression that the impurities give rise to the band of propagating excitations inside the {\it restrahl} of the original crystal, with boundaries given by 
\begin{eqnarray}
\omega _{il}^{2} = \Omega _{1}^{2} - p 
\left( 
\frac { \frac {2}{3} d_{0}^{2}  (\Omega _{1}^{2}-\omega ^{2}) - \frac {2}{3} d_{1}^{2}  (\Omega _{0}^{2}-\omega ^{2})}              
{ \Omega _{0}^{(LO)2} - \Omega _{1}^{2} } \right), 
\nonumber \\
\omega _{iu}^{2} = \Omega _{1}^{2}+p \left( \frac 
{\frac {\epsilon _{\infty }}{3}d_{1}^{2}(\omega ^{2}-\Omega _{0}^{2})-\frac {\epsilon _{\infty }}{3}d_{0}^{2}(\omega ^{2}-\Omega _{1}^{2})}{  \Omega _{1}^{2} - \Omega _{0}^{(TO)2} } \right) .
\label{Boundaries}
\end{eqnarray}
The width of the band (in terms of squared frequencies) is linearly proportional to the concentration
\begin{equation}
\Delta_{imp}^{2} \simeq d_{1}^{2}  (\Omega_{1}^{2}-\Omega_{0}^{2}) \frac
{\frac{ 4 \epsilon_{\infty}}{3} d_{0}^{2} -\frac {\epsilon_{\infty} +2 }{3} (\Omega_{1}^{2}-\Omega_{0}^{2})} {\left( \Omega _{0}^{(LO)2} - \Omega _{1}^{2} \right) \left( \Omega _{1}^{2} - \Omega _{0}^{(TO)2}\right) } \cdot p \ >0 \ ,
\label{Band_width}
\end{equation}
provided that the impurity frequency $\Omega_{1}$ is not too close to $\Omega _{0}^{(LO)}$ such that 
\begin{equation}
\Omega _{0}^{(LO)\ 2}-\Omega_{1}^{2}<\frac {2}{3}d_{0}d_{1}\sqrt{p}.
\label{condition}
\end{equation}
The lower band boundary, given by Eq. (\ref{Boundaries}), corresponds to a dispersionless longitudinal polariton branch
\begin{equation}
\omega _{i}^{(LO)}({\bf k})=\omega _{il},
\label{LO_dispersion}
\end{equation}
while the upper one is the short wave limit of the branch of transversal excitations. An approximate dispersion law for these excitations can be obtained from Eq. (\ref {transv_k}) if one substitutes $\omega=\Omega_1$ everywhere except in terms containing impurity polariton band boundaries:
\begin{equation}
\omega _{i}^{(TO)}({\bf k})=\omega _{il}+\delta _{imp}\frac {k^{2}l_{0}^{2}}{1+k^{2}l_{0}^{2}}.
\label{TO_dispersion}
\end{equation}
 Parameter $l_0$ in this equation is the  localization length of a single local polariton with the frequency equal to $\Omega_1$:
\begin{equation}
l_{0}^{-1}=\left( \frac{\Omega_1 ^{2} }{c^{2}}\epsilon _{\infty } \frac{\Omega _{0}^{(LO)2}-\Omega_1 ^{2}}{\Omega_1 ^{2}-\Omega _{0}^{(TO)2}}\right) ^{1/2}.
\label{loclength}
\end{equation}

Excitations described by Eqs. (\ref{LO_dispersion}) and (\ref{TO_dispersion}) demonstrate a number of peculiarities. First, one can note that the mutual positions of longitudinal and transverse modes are reversed compared to the regular polaritons: the longitudinal mode has lower frequency than the transverse one. However, if one takes the original polariton branches of the host crystal into consideration, the normal sequence of transverse  and longitudinal modes is restored: the host transverse  polariton branch is followed by the impurity longitudinal mode, which is followed by the impurity transverse mode. The last modes in the sequence are the LO mode and upper transverse  polariton branch of the host. Second, the single transverse impurity polariton mode combines properties of lower and upper regular polariton modes. Indeed, at $k=0$ this mode becomes degenerate with the impurity longitudinal mode akin to the upper branch of regular polaritons. At the same time the short wave limit $k\rightarrow 0$ of the same mode corresponds to the TO frequency of the electrostatic approximation, and sets the upper boundary of  the propagating band similar to the regular lower polariton branch.  
\begin{figure}
\centering
\epsfxsize=3in \epsfbox{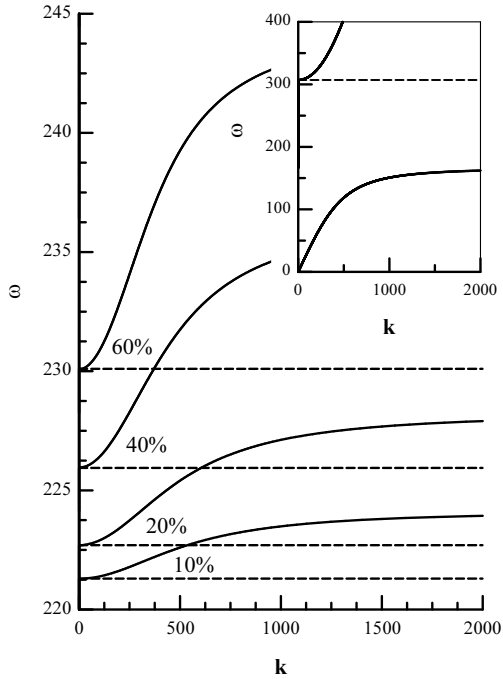}
\caption{Dispersion curves of the transverse  (solid) and longitudinal (dashed) optic defect modes in the {\it restrahl} for four different concentrations. The insert depicts the dispersion curves of the pure crystal. Frequency and wavenumber are given in the same units, $cm^{-1}$.}
\end{figure}

The dispersion curves of the transverse  excitations, obtained from general equation (\ref{transv_k}) for several concentrations, are shown in Fig. 1. The similar dispersions were observed experimentally by means of Raman spectroscopy in mixed crystal $Ga/InP$ in Ref.\onlinecite{Livescu}. Originally, the interpretation of these observations was based upon a phenomenological dielectric function with multiple resonances, while it seems clear now that they provide a solid support for the concept of impurity polaritons.

From the dispersion law described by Eq. (\ref{TO_dispersion}) one can obtain an expression for the group velocity of the respective excitations:
\begin{equation}
\frac {v_{g}(\omega )}{c}=\frac {l_{0}}{\lambda }\frac{\left( \omega ^{2}-\omega _{il}^{2} \right)
^{1/2}\left( \omega_{iu}^{2} -\omega ^{2}\right) ^{3/2}}{\omega ^{2}\Delta _{imp}^{2} },  
\label{v_group}
\end{equation}
which  reaches its maximum value of 
\begin{equation}
\frac {v_{g}(\omega _{max})}{c}=\frac {l_{0}}{\lambda }\frac {3^{3/2}}{8}\frac {\delta _{imp}}{\omega }\sim \frac {\delta _{imp}}{\omega },  
\label{v_group_max}
\end{equation}
at 
\[\omega ^{2}_{max}=\omega _{il}^{2}+1/4\Delta _{imp}^2.
\]
We introduced here a new parameter, $\delta _{imp}$, characterizing the width of the impurity polariton band in terms of frequencies themselves: $\delta _{imp}\simeq \Delta _{imp}^{2}/2 \omega$. The group velocity is linear in concentration and is significantly smaller than the speed of light in vacuum. This is an interesting result showing that propagation of light through mixed crystals can be significantly (and controllably) slowed down. The smallness of the group velocity is also reflected in the flatness of the dispersion curves presented in Fig. 1, as well as in experimental results of Ref.\onlinecite{Livescu}

The density of polariton states in the impurity band (IPDOS) can also be obtained from the dispersion equation, Eq. (\ref{TO_dispersion}),
\begin{equation}
D(\omega ) \simeq  \frac {1}{2 \pi ^{2}l_{0}^{3}} \omega  \Delta _{imp}^{2}\frac{\left( \omega ^{2}-\omega _{il}^{2} \right)
^{1/2}}{\left( \omega_{iu}^{2} -\omega ^{2}\right) ^{5/2}}.  
\label{Dos_small_p}
\end{equation}
At the low-frequency boundary of the band this DOS reproduces the singular behavior characteristics for the lower polariton band of  pure polar crystals, while at the high frequency edge it resembles the behavior of the regular upper polariton band. At the center of the impurity band IPDOS can be expressed in compact form in terms of the localization length $l_0$ and the linear width of the band $\delta _{imp}$:
\begin{equation}
D(\omega _{c}) \simeq  \frac {1}{ \pi ^{2}l_{0}^{3}\delta _{imp}  }.
\label{Dos_band_center}
\end{equation}
The dependence of IPDOS upon concentration is different at the edges of the band and at the center. The boundary of IPDOS depends linearly upon $p$, and tends to zero when $p$ decreases. At the band center IPDOS is inverse proportional to $p$, and goes to infinity when $p\rightarrow 0$. Such a behavior has a simple meaning -- it describes the collapse of the band into a single local state with an infinite density.

The linear in concentration approximation for the impurity band fails when the defect frequency $\Omega _{1}$ falls close to the band edge, $\Omega _{0}^{(LO)}$, of the pure crystal, so that Eq. (\ref{condition}) does not hold anymore. In this case, one can obtain approximate expressions for characteristics of the impurity band using an expansion of Eq. (\ref {Epsilon}) in powers of $\Omega _{0}^{(LO)}-\Omega _{1}$. The zeroth order in this parameter leads to a square root dependence of the band width upon concentration:
\begin{equation}
\Delta_{imp}^{2} \simeq \frac {2d_{0} d_{1} }{3} \sqrt {p}.
\label{BW_sqr}
\end{equation}
It is interesting to note that this situation was probably observed in one-two-mode mixed crystal $Ga/InAs$.\cite{Groenen} In that system at small concentration of $In$ the frequencies $\omega _{il}(p)$ and $\Omega ^{(LO)}(p)$ could not be fit with linear in $p$ expressions, while $\omega _{iu}(p)$ remained linear. This is exactly what our model predicts in the case when $\Omega _{1}$ approaches $\Omega _{0}^{(LO)}$:
\begin{eqnarray}
\omega _{il}^{2} &=&\Omega _{0}^{(LO)2}-\frac{2d_{0}d_{1}}{3}\sqrt{p},\nonumber \\
\omega _{iu}^{2} &=&\Omega _{0}^{(LO)2}
+\frac{2}{3}\frac{\epsilon _{\infty }}{2+\epsilon_{\infty}}d_{1}^2p,
\label{SqrtBoundaries} \\
\Omega ^{(LO)\ 2} &=&\Omega _{0}^{(LO)\
2}+\frac{2d_{0}d_{1}}{3}\sqrt{p}.
\nonumber
\end{eqnarray}
DOS in this case is given by the expression
\begin{equation}
D(\omega ) \simeq  \frac {\Omega _{0}^{(LO)}}{2 \pi ^{2}\lambda ^{3}} \ \frac {(\omega ^{2}-\Omega _{0}^{(LO)\ 2})^{2}+\Delta _{imp}^{4}}{\left( \displaystyle{\frac {2+\epsilon _{\infty }}{3\epsilon _{\infty}}} d_{0}^{2}\right) ^{3/2} } \    
\frac{\left( \omega ^{2}-\omega _{il}^{2} \right) ^{1/2}}{\left( \omega_{iu}^{2} -\omega ^{2}\right) ^{5/2}},  
\label{DosSqrtp}
\end{equation}
which has the regular polariton behavior at the boundaries similar to one described by Eq. (\ref {Dos_small_p}), but with a different pre-factor. This pre-factor significantly modifies IPDOS at the center of the band,
\begin{equation}
D(\omega _{c}) \simeq  
\frac {3^{1/2}5}{ \pi ^{2}}
\frac {1}{\lambda ^{3} \Omega _{0}^{(LO)}} 
\left(  \frac {\Omega _{0}^{(LO)\ 2}}{\frac {2+\epsilon _{\infty }}{3\epsilon _{\infty}}d_{0}^{2}}\right) ^{3/2} 
\left( \frac {\delta _{imp}}{\Omega _{0}^{(LO)}} \right) ^{1/2},
\label{DosSqrtpCenter}
\end{equation}
which is now proportional to $p^{1/4}$. At very small concentrations, however, this dependence will change over to the one described by Eq. (\ref{Dos_band_center}) as the condition given by Eq. (\ref{condition}) becomes valid again.

The dispersion relation for the impurity TO branch becomes in this limit more complicated: 
\begin{eqnarray}
\omega _{i}^{(TO)\ 2}({\bf k}) &=& \omega _{il}^{2}+\frac {1}{2}\left[ 2\Delta _{imp}^{2}+k^{2}\lambda ^{2}\left( \frac {2+\epsilon _{\infty }}{3\epsilon _{\infty}} d_{0}^{2} \right) \right.
\nonumber \\
&-& \left. \sqrt {k^{4}\lambda ^{4}\left( \frac {2+\epsilon _{\infty }}{3\epsilon _{\infty}} d_{0}^{2} \right)^{2}+4\Delta _{imp}^{4}}
\right],
\label{sqrtpTO_disp}
\end{eqnarray}
while it still has the same limits $\omega _{i}^{(TO)}({\bf k}) - \omega _{il} \sim k^{2}$ for small $k$ and $\omega _{i}^{(TO)}({\bf k}) = \omega _{iu}$ for large $k$. The crossover parameter, however, is no longer the localization length $l_0$, but rather a vacuum wavelength of light at frequency $\Omega^{LO}$: $\lambda=\Omega^{LO}/c$. The effective mass of the branch at the longwave boundary is much larger than in Eq. (\ref{TO_dispersion}) and does not depend upon concentration. 
The LO frequency of the impurity polariton branch does not show any dispersion:
\begin{equation}
\omega _{i}^{(LO)}({\bf k})=\omega _{il},
\label{sqrtpLO_disp}
\end{equation}
similar to the situation considered before, for $k^{2}>\Delta _{imp}^2/(\lambda ^2 d _0^2)$ two branches of the impurity band run almost parallel showing a very small dispersion.

The group velocity at the boundaries approaches zero at the same rate as before, but at the center of the band, the dependence upon the concentration is different:
\begin{equation}
\frac {v_{g}(\omega _{c})}{c}\simeq \frac {3^{1/2}}{5}\left(  \frac {2+\epsilon _{\infty }}{3\epsilon _{\infty}}\frac {d_{0}^{2}}{\Omega _{0}^{(LO)\ 2}}\right) ^{1/2}\left( \frac {\delta _{imp}}{\Omega _{0}^{(LO)}} \right) ^{1/2} .
\label{vg_sqrtp}
\end{equation}
It is much larger than the respective quantity, when the linear in $p$ expansion is valid, and  increases much faster $\sim p^{1/4}$ with the concentration.

\subsection {Evolution of the polariton impurity band boundaries with composition parameter}

Further increase of concentration leads to more complicated dependencies  of the bandwidth, DOS, {\it etc.} on concentration and other parameters. In this subsection we shall focus upon concentration dependencies of band boundaries, which are experimentally identified with TO and LO phonon frequencies of the electrostatic approximation. These dependencies were extensively studied experimentally, as we have already discussed in the Introduction, and the objective of this subsection is to show how the concept of impurity induced polaritons provides a simple and physically transparent explanation for the weak mode in one-mode crystals, and for one-two-mode behavior.  Analysis of poles and zeroes of the dielectric function (\ref{Epsilon}) shows that the evolution of the modes with concentration  is determined in our model by a relative position of the characteristic frequencies $\Omega _0, \ \Omega _1, \ \Omega _0^{TO}, \ \Omega _0^{LO}, \ \Omega _1^{TO}$, and $\Omega _1^{LO}$. The first pair of these frequencies are the initial phonon frequency of the host crystal, and the local polariton mode of the impurity, respectively. The electrostatic interaction, and  the local field changes $\Omega_0$ to actual TO and LO host frequencies $\Omega _{0}^{TO}$ and $\Omega _{0}^{LO}$. The last pair of the frequencies, $\Omega _{1}^{TO}$ and $\Omega _{1}^{LO}$, corresponds to TO and LO modes of the crystal made up of impurity atoms only. The local field, which induces the difference between initial frequencies, $\Omega _{0,1}$, and actual TO, LO frequencies, is very important. In pure crystals (made of initial host atoms or initial impurities) the relation between $\Omega _{0,1}$ and $\Omega _{0,1}^{TO}$ and $\Omega _{0,1}^{LO}$,
\begin{equation}
(2+\epsilon _{\infty })\Omega _{0,1}^{2}=2\Omega _{0,1}^{(TO)\ 2}+\epsilon _{\infty }\Omega _{0,1}^{(LO)\ 2},
\label{Omegas}
\end{equation}
was originally derived in Ref.\onlinecite{Born-Huang}, and its importance was stressed in Ref.\onlinecite{Bonneville}. The same relation  exists in our model as well. 
\begin{figure}
\centering
\epsfxsize=4in \epsfbox{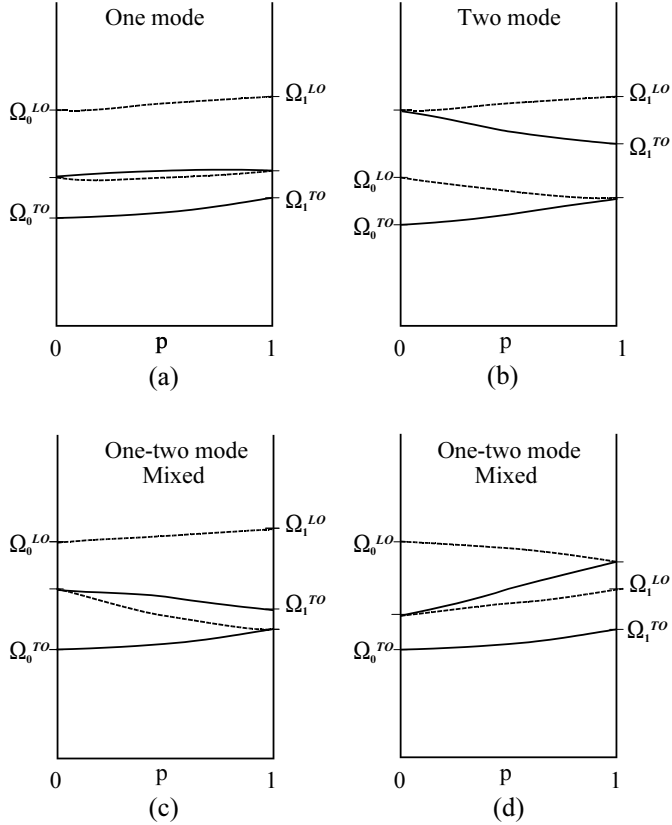}
\vspace{-.4in}
\caption{Four different types of the evolution of the transverse  (solid lines) and longitudinal (dashed lines) optic mode frequencies of the mixed polar crystals with composition parameter. The leftmost and rightmost composition parameters correspond to pure end crystals.}
\end{figure}
At small concentrations, $p\ll 1$, the initial impurity related frequency $\Omega_1$ is not renormalized by local-field corrections  because this renormalization is caused by the interaction with like atoms. At small concentrations an impurity atom is mostly surrounded by atoms of the host crystal, and the renormalization does not occur. Similarly, for $1-p\ll 1$, the local field turns $\Omega_1$ into  $\Omega _{1}^{TO}$ and $\Omega _{1}^{LO}$, while leaving $\Omega_0$ unchanged as a characteristic frequency of former host atoms. 
Eqs. (\ref {Boundaries}) demonstrate that at small $p$ the polariton impurity band inside the {\it restrahlen} of the host  arises  when $\Omega _{1}$ falls inside the host {\it restrahlen}, see Fig. 2a,c,d. 
For $1-p \ll 1$, host and impurity atoms exchange their roles. In this case, should $\Omega _{0}$ falls in between $\Omega _{1}^{(TO)}$ and $\Omega _{1}^{(LO)}$, it gives rise to the impurity polariton band induced now by the ``host'' atoms. Hence, if both $\Omega _{0}^{(TO)}<\Omega _{1}<\Omega _{0}^{(LO)}$ and $\Omega _{1}^{(TO)}<\Omega _{0}<\Omega _{1}^{(LO)}$ conditions are satisfied at the same time, then the polariton impurity bands exist in both $p\rightarrow 0$ and $p\rightarrow 1$ limits. This situation is shown in Fig. 2a. This is one-mode behavior with a ``weak" mode. One observes strong TO and LO modes of the original crystal smoothly evolving into those of the end crystal, while inside the {\it restrahlen} a weak additional TO and LO impurity polariton modes arise with vanishing TO-LO splitting at both ends of the impurity range.  This peculiar behavior occurs due to the interplay between the external electric field and the polarization, affecting the local field (\ref {E_loc}). Experimentally, this type might be realized in the materials where the polariton gap is sufficiently wide, which may be the case for many alkali halides. However, because of the weakness of these modes they are vulnerable to any kind of dissipation, as we shall see in the next section. This fact can explain the absence of the defect mode in ``classical'' (no weak mode inside the {\it restrahlen}) one-mode mixed crystals. At the same time, we can relate the features presented in Fig. 2a to the weak mode observed in the  $Ba/SrF_{2},Ca/SrF_{2}$ and some other one-mode crystals (see Discussion). Previously, in order to reproduce the weak feature observed in spectra of these crystals, one had to use models with tens of fitting parameters (see Refs. \onlinecite{Sievers,Taylor-review} and references therein). The concept of the polariton impurity band, presented here, gives a transparent and quite general explanation of this type of spectra.

If the polariton impurity band exists only in one limit, for example, $\Omega _{0}^{(TO)}<\Omega _{1}<\Omega _{0}^{(LO)}$  for small $p$, but $\Omega _{0}$ lies outside the interval $(\Omega _{1}^{(TO)},\Omega _{1}^{(LO)})$ when $p\sim 1$, our model predicts the  one-two-mode behavior (Fig. 2c,d). Two types of one-two-mode mixed crystals appear depending upon whether $\Omega _{0}$ falls below the interval $(\Omega _{1}^{(TO)},\Omega _{1}^{(LO)})$ (Fig. 2c), or above it (Fig. 2d). In the first case, the lower mode weakens with the concentration, while in the second case, the upper one does. At $p \simeq 0$ the situation remains qualitatively the same as in the one-mode case. At $p \rightarrow 1$, the splitting between new modes does not vanish, and they form the new {\it restrahlen} of the pure crystal at $p=1$. Strictly speaking, in this limit we cannot justify our model, since there are no local modes at this end of the concentration range. Quantitatively however, we can still explain the one-two-mode behavior as a result of the presence of the local polaritons at small $p$, and the absence of them at $p$ close to one.
If in neither limit the impurity polariton band appears inside the gap ($\Omega _{1}\not\subset(\Omega _{0}^{(TO)},\Omega _{0}^{(LO)})$ and $\Omega _{0}\not\subset(\Omega _{1}^{(TO)},\Omega _{1}^{(LO)})$), our approach is not applicable at any concentration. Although the absence of the impurity bands at end concentrations, in our opinion, suggests two-mode behavior (Fig. 2d).

\section {Transmittance and reflectance of EM waves from mixed polar crystals}

Having obtained the effective dielectric function of the mixed polar crystal, one can study their reflection and transmission spectra for a number of different geometries. Among them, the normal reflection spectrum from a semi-infinite sample and normal transmission and reflection from a slab of finite dimensions are of particular interest since they are a primary source of information about optical  properties of  crystals studied in numerous experiments.\cite {Sievers,Taylor-review}

\subsection {Reflection spectrum of semi-infinite mixed polar crystals}

In this section we phenomenologically incorporate damping in the dielectric function (\ref{Epsilon}) substituting $\omega+i\gamma$ for $\omega$ in the resonance terms of Eq. (\ref{Epsilon}). As we already discussed it is sufficient for our qualitative purposes to assume that the damping parameter is the same for both host and impurity related modes. We can note, however, that as numerical calculations demonstrated, properties of each mode are determined primarily by its own  relaxation parameters, and effects due to damping in the other subsystem are usually negligible. This means that, in principle, it is possible to experimentally determine relaxation parameters for each of the participating oscillators independently. For our calculations we chose the value of $\gamma$ such that $\gamma /\Omega _{0} \sim 0.1$, which is a rather conservative estimate. For typical polar crystals the relaxation parameter ranges from $\gamma /\Omega _{0} $ less than $0.01$ to $0.1$, and since the relaxation in the impurity subsystem due to coupling to LO phonons inside the {\it restrahlen} was estimated in Section III as less than $0.1$, our choice for this parameter seems quite reasonable.

 After accounting for the dumping, the effective dielectric function (\ref {Epsilon}) becomes
\begin{equation}
\epsilon (\omega )={\epsilon}_{\infty}  \frac
{1+\displaystyle{\frac{2}{3}} \left[ (1-p)\displaystyle{\frac{d_{0}^{2}}{\Omega _{0}^{2}-\omega ^{2}+2i\gamma \omega }}+p\displaystyle{\frac{d_{1}^{2}}{\Omega _{1}^{2}-\omega ^{2}+2i\gamma \omega }} \right]}
{1-\displaystyle{\frac{{\epsilon}_{\infty}}{3}} \left[ (1-p)\displaystyle{\frac{d_{0}^{2}}{\Omega _{0}^{2}-\omega ^{2}+2i\gamma \omega }}+p\displaystyle{\frac{d_{1}^{2}}{\Omega _{1}^{2}-\omega ^{2}+2i\gamma \omega }} \right]}.
\label{Epsilon_dump}
\end{equation}
For normal incidence, the reflection coefficient from the semi infinite crystal is equal to 
\begin{equation}
R(\omega )=\left| \frac {1-\epsilon ^{1/2}(\omega)}{1+\epsilon ^{1/2}(\omega)}\right| ^{2}.
\label{R_semiinf}
\end{equation}

In the absence of damping, the original dielectric function (\ref{Epsilon}) has a peculiar property, which is specific only for the model with the impurity polariton band within the {\it restrahlen} of the host. Inside this band, the dielectric function goes from zero to infinity, and hence, necessarily   passes through unity. At a frequency where this happens, the reflection coefficient must become zero, since for this frequency the medium becomes transparent. At small concentrations, this frequency, $\omega _{T=1}$, is determined by the equation:
\begin{equation}
\omega _{T=1}^{2}=\frac {\lambda_1 ^{2}\omega _{il}^{2}+l_{0}^{2}\omega _{iu}^{2}}{\lambda_1 ^{2}+l_{0}^{2}},
\label{omega_one}
\end{equation}
and since $l_{0}\lesssim \lambda_1=\Omega_{1}/c$ it  lies slightly off the center of the band, closer to $\omega _{il}$.
When relaxation is accounted for, the zero of reflection is not reached, but the reflection still can have a minimum at a certain frequency. The magnitude of reflection at this frequency is determined by relaxation, and can be used for independent measurements of the latter. 

We used typical values of the parameters in Eq. (\ref {Epsilon_dump}) to plot normal reflection spectra for semi-infinite mixed crystals. The optic frequencies $\Omega _{0,1}^{(TO,LO)}$ of such systems are about a few hundred $cm^{-1}$, the widths of their polariton gaps range from $\sim 10\% $ for III/V group polar crystals, up to $\sim 30\% $ for alkali halides. The high limit dielectric constant lies within the range of $\sim 3 - 5$.  Fig. 3 presents three graphs corresponding to three different types of spectra, which can be described within the model of impurity polaritons. Each graph shows curves obtained for $ p=0\%,\ 25\%,\ 50\%,\ 75\%,\ 100\% $. 
\begin{figure}
\centering
\epsfxsize=4in \epsfbox{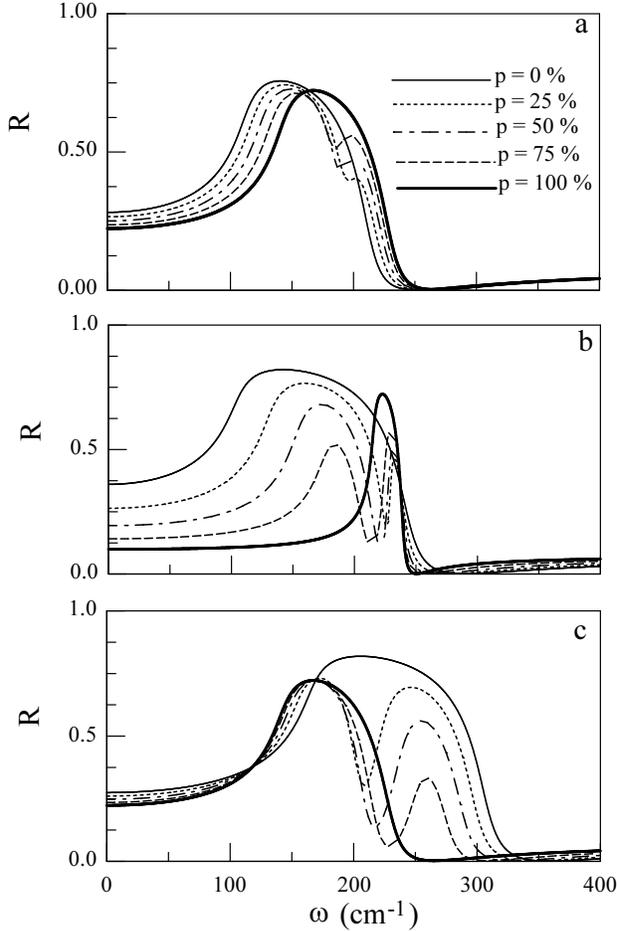}
\caption{Normal reflection spectra from semi infinite mixed crystals. Three graphs correspond to one-mode (graph (a) in Fig. 2) and two types of one-two-mode (graphs (c) and (d) in Fig. 2) behaviors. The parameters used to generate these graphs (including absorption) are typical parameters for polar crystals.}
\end{figure}
Fig. 3a depicts the reflection spectrum of the one-mode crystal. As it can be seen from the plot, for all concentrations there is one dominating absorption band. Nevertheless, for intermediate concentrations one can notice a weak mode inside this band. This mode is quite weak in accord with the discussion of the previous section, and can be smoothed away by the absorption. Whether this mode will be observed in a concrete material depends upon the interplay of several parameters. At the same time, since we have used realistic values of parameters characteristic for one-mode group of mixed crystals, our calculations show that the impurity polaritons can indeed be used to explain this feature of one-mode type spectra.
Reflection spectra corresponding to one-two-mode behavior are shown in Figs. 3b and 3c. There are two types of such spectra, which are very much alike. At small $p$, the spectra look akin to the one-mode type, but with an increase of the concentration two modes appear, with one-mode growing stronger and the second one diminishing. Spectra 3b and 3c correspond to phase diagrams shown in Figs. 2c and 2d, respectively. It is seen that the presence of damping does not prevent one-two-mode behavior predicted by our model to be observed in reflectance experiments.

\subsection {Slab of finite dimensions}

In this subsection we consider normal reflection and transmission spectra of a mixed crystal slab. The width of the slab is assumed to be much greater than the average distance between the impurities, so that our averaging procedure can be applied. At the same time we do not consider samples thicker than several wavelength, so that the damping does not suppress transmission completely.

The transmission coefficient through a slab of width $L$ for normal incidence is given by
\begin{equation}
T(\omega )=\left| 1-{\displaystyle \frac {(\epsilon ^{1/2}-1)^2\cos\left( \frac{\omega }{c}\epsilon ^{1/2}L\right) }{ 2\epsilon ^{1/2}\cos\left( \frac{\omega }{c}\epsilon ^{1/2}L\right)-i(\epsilon +1)\sin\left( \frac{\omega }{c}\epsilon ^{1/2}L\right) }}\right| ^{2}.
\label{T_slab}
\end{equation}
In the absence of damping, the transmission coefficient turns to unity when either $\epsilon =1$ or $\omega \epsilon ^{1/2}L/c=\pi (m+1/2)$, where $m$ is an integer. The first case corresponds to the frequency determined by Eq. (\ref{omega_one}), when the medium becomes optically transparent, while the second condition corresponds to the usual geometrical resonances. 
\begin{figure}
\centering
\epsfxsize=4in \epsfbox{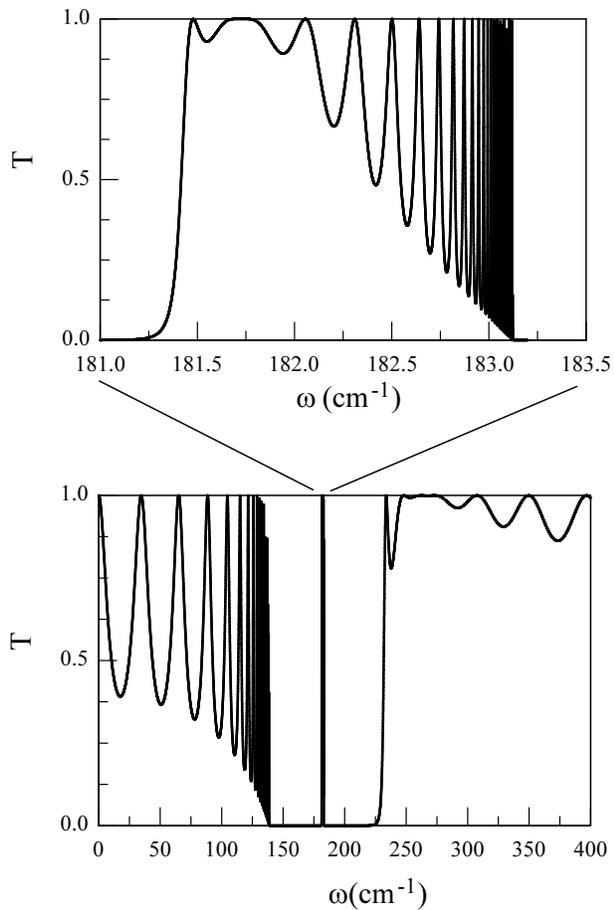}
\caption{Transmittance spectrum through a slab of one wavelength width with typical parameters, plotted as a function of frequency. The lower graph shows the full frequency range, while the upper one is restricted to the polariton defect band at $p=10\%$.}
\end{figure}

The width of geometrical resonances decreases as $1/\epsilon (\omega )$ with $\omega \rightarrow \omega _{iu}$ because of the divergence of the dielectric function at this frequency. Fig. 4 depicts the transmission spectrum of the $10\%$ impure slab of the width $L=\lambda$ in the absence of absorption, where the lower plot shows the broader frequency interval covering the entire {\it restrahlen} of the host crystal. One can see the impurity induced polariton band inside the forbidden gap at the frequency $\sim 180 \ cm^{-1}$.  The wide flat-top resonance at $181.7 \ cm^{-1}$ corresponds to the frequency (\ref{omega_one}), where the dielectric function becomes one, and all the other peaks represent geometrical resonances. 
\begin{figure}
\centering
\epsfxsize=2.5in \epsfbox{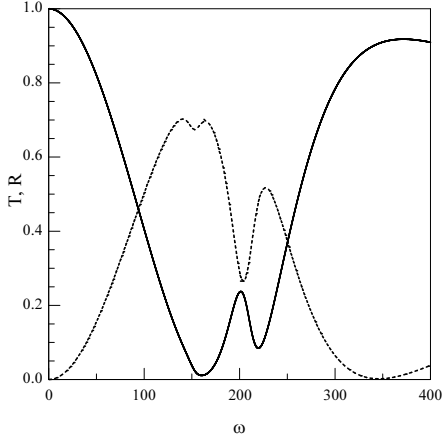}
\vspace{-.6in}
\caption{Transmission (solid) and reflection (dashed) coefficients of a thin mixed crystal at $p=20\%$ in the presence of absorption.}
\end{figure}

Different natures of the resonances affect their response to damping. The narrow comb of geometrical resonances washes out first even when damping is relatively small. This occurs due to the fact that the geometrical resonances are in essence standing waves that experience multiple reflections and therefore are strongly damped. The $\epsilon =1$ resonance, to the contrary, is much less affected by the relaxation because polaritons pass through the sample only once. 
Fig. 5 depicts the transmission (solid line) and reflection (dashed line) coefficients for a narrow slab with $20\%$ impurities and the same set of parameters, which was used to generate Fig. 3c. It is clearly seen, that even despite of high absorption rate $\gamma /\Omega _{0}=0.1$, the peak in the transmission coefficient survives, and can be associated with the minimum in reflection. It allows us to suggest that reflection anti-resonance observed in many systems can be associated not only with the impurity induced absorption, but also with the impurity induced transmission. A similar transmission maximum was observed experimentally in $CuCl$,\cite{CuClTransm} where the role of defects was played by some $Cu$ atoms occupying off-center positions. The maximum in the transmission is direct evidence of the impurity-induced polariton band. It would be of great interest to carry out transmission measurements for different groups of mixed crystals in order to verify predictions of the current paper. We believe that time-resolved measurements could also provide important information, particularly regarding group velocities of these excitations.

\section {Scattering in the impurity-induced polariton band}

In previous sections we considered optical properties of  mixed polar crystals neglecting fluctuations of concentration function $p({\bf r})$. Results obtained for the one-dimensional model considered in our previous paper, Ref.\onlinecite{Lyapunov}, suggests that this approximation is a zero-order term in the series expansion in terms of the parameter $l/l_0\ll 1$, where $l$ is an average distance between defects and $l_0$ is the localization length of the single local polariton. In this section we take into account fluctuations of concentration in the first non-vanishing order in terms of the parameter $\delta p = p({\bf r})-p$, and show that the actual small parameter of this approximation is indeed $l/l_0\ll 1$. The main result of this section is to demonstrate that the scattering of impurity polaritons induced by concentration fluctuations is decreasing with an increase of the average concentration, and is actually negligible on the background of absorption. This fact provides a justification for the results  obtained in the previous sections of the paper, where these fluctuations are neglected.

We shall employ the Green's function formalism in order to calculate the scattering length of the impurity-induced polaritons due to concentration fluctuations. The Green's function of the inhomogeneous dielectric medium can be presented as
\begin{eqnarray} 
\left[ G _{\alpha \gamma }^{(0)\ -1}({\bf r},{\bf r'})-  \frac {\omega ^2}{c^2}\left[ \epsilon (\omega ,p({\bf r})) -\epsilon (\omega ,p)\right]    \delta _{\alpha \gamma }\right] G _{\gamma \beta }({\bf r},{\bf r'}) 
\nonumber \\
=\delta _{\alpha \beta }\delta ({\bf r}-{\bf r'}),
\label{G_eq}
\end{eqnarray}
where $G _{\alpha \gamma }^{(0)}$ is the Green's function of the system with homogeneous dielectric function $\epsilon (\omega ,p)$ given by Eq. (\ref{Epsilon}). This zeroth order Green's function can be written down in $k$-space in terms of the projection operators   ${\displaystyle \hat {\bf e}^{\perp }_{\alpha \beta }=\delta _{\alpha \beta }- k_{\alpha }k_{\beta }/k^{2}}$ and ${\displaystyle \hat {\bf e}^{\parallel }_{\alpha \beta }=k_{\alpha }k_{\beta }/k^{2}}$ as
\begin{equation} 
G _{\alpha \beta }^{(0)}({\bf k})=\frac {\hat {\bf e}^{\perp }_{\alpha \beta }}{k^{2}-k^{2}_{0}}-\frac {\hat {\bf e}^{\parallel }_{\alpha \beta }}{k^{2}_{0}}.
\label{G_zero}
\end{equation}
where $k_{0}^{2}(\omega )= \omega ^2 \epsilon (\omega )/c^2$. 
Since we search for the leading corrections to the Greens' function of the system, we expand $\epsilon (\omega ,p({\bf r}))$ in terms of $\delta p({\bf r})$ and keep only the linear term
$$
\epsilon (\omega ,p({\bf r})) -\epsilon (\omega ,p)=\left. \frac {\omega ^2}{c^2}\frac {\delta \epsilon (\omega ,p+\delta p({\bf r}))}{\delta p({\bf r})}  \right| _{\delta p({\bf r})\equiv 0}\cdot \delta p({\bf r})$$
$$
=\kappa ^{2}(\omega ,p)\cdot \delta p({\bf r}).
$$
Statistical properties of the random function $\delta p({\bf r})$ can be described in this approximation by its momenta up to the second order:
\begin{eqnarray} 
\left\langle {\delta p({\bf r})}\right\rangle &\equiv & 0 \ ,
\nonumber \\
\left\langle {\delta p({\bf r})\delta p({\bf r'})}\right\rangle &=& K(\left| {\bf r}-{\bf r'}\right| ).
\label{Correlator}
\end{eqnarray}
The second order correlator depends only upon the distance between two points in space, since the system is assumed to be homogeneous and isotropic on average. After standard transformations to $k$-representation one obtains
\begin{eqnarray} 
G _{\alpha \beta }^{-1}({\bf k}) &=& (k^{2}-k_{0}^{2}) \hat {\bf e}^{\perp }_{\alpha \beta }-k_{0}^{2} \hat {\bf e}^{\parallel }_{\alpha \beta }-
\\
\nonumber
&-&\kappa ^{4}\int \frac {d^{3}k'}{(2\pi )^{3}}\left( \frac {\hat {\bf e}^{\perp }_{\alpha \beta }}{k'^{2}-k^{2}_{0}}-\frac {\hat {\bf e}^{\parallel }_{\alpha \beta }}{k^{2}_{0}}\right) S({\bf k}-{\bf k'})
\label{G_inv} \\
&=& (k^{2}-k_{0}^{2}-\Sigma ^{\perp}(k)) \hat {\bf e}^{\perp }_{\alpha \beta }-(k_{0}^{2}+\Sigma ^{\parallel }(k))\hat {\bf e}^{\parallel }_{\alpha \beta },
\nonumber
\end{eqnarray}
where $S({\bf k})$ is the Fourier transform of the correlator (\ref{Correlator}). The new Green's function can be expressed using transverse  and longitudinal  mass operators $\Sigma ^{\perp}(k),\Sigma ^{\parallel}(k)$:
\begin{equation} 
G _{\alpha \beta }({\bf k})=\frac{\hat {\bf e}^{\perp }_{\alpha \beta }}{k^{2}-k_{0}^{2}-\Sigma ^{\perp}(k)}-\frac {\hat {\bf e}^{\parallel }_{\alpha \beta }}{k_{0}^{2}+\Sigma ^{\parallel }(k)}.
\label{G}
\end{equation}
Real parts of the mass operators determine the renormalization of the spectrum, and are neglected below. We shall only consider imaginary parts of the transverse mass operators, which  determine the scattering length (mean-free-path) of the transverse modes, and in the lowest approximation read as
 \begin{equation} 
Im\ \Sigma ^{\perp}(k_{0}(\omega ),\omega )=i\frac {\kappa ^{4}}{2\pi }k_{0}\int S\left( 2k_{0} \sin \frac {\theta }{2}\right) \ d \cos\theta ,
\label{Sigma_im}
\end{equation}
where $\theta $ is scattering angle between ${\bf k}$ and ${\bf k'}$. The explicit form of the integral scattering cross section depends on the particular choice of the correlator (\ref {Correlator}). In our case however, when the wavelength of the considered excitations is assumed to be much greater than the characteristic size of the inhomogeneities, the difference between different choices of the correlation function is mostly reduced to a numerical factor of the order of unity. One can choose the correlator, for example, in the standard Gaussian form:\cite{Ishimaru} 
\begin{equation} 
K(\left| {\bf r}-{\bf r'}\right| )=\left\langle \delta p^2\right\rangle \ e^{-\left| {\bf r}-{\bf r'}\right|^2/l_c^2}
\label{F_r}
\end{equation}
with its respective Fourier transform
\begin{equation} 
S(k)={\displaystyle \frac {\left\langle {\delta p^{2}}\right\rangle \pi ^{3/2}l_{c}^{3}}{8}\ }e^{ -k^{2}l_{c}^{2}/4}.
\label{F_k}
\end{equation}
In the spirit of our general approach,  the concentration fluctuations must be considered with regard to the smoothing volume $\delta V,$ so that the respective variance $\left\langle {\delta p^{2}}\right\rangle $ is calculated using the Poisson distribution of independent impurities inside the volume $\delta V$:
\begin{equation} 
\left\langle {\delta p^{2}}\right\rangle = \frac {p(1-p)}{N(\delta V)},
\label{Delta_p3d}
\end{equation}
The correlation length $l_{c}$, then should be identified with the $l_{\delta V}$-smoothing length we employed in Eqs. (\ref{Smoothing1}) and (\ref{Smoothing2}). However, as will be seen below, the same results can be obtained if one considers initial distribution of discrete impurities and chooses the interatomic distance $a$ as the correlation length. In any case,  $k_{0}l_{c}\ll 1$, and  the expression for the complex wave number $k(\omega )$ determined from the pole of the respective Green's function takes the form:
\begin{equation} 
k(\omega )=k_{0}(\omega )+i{\displaystyle \frac {\sqrt {\pi }}{16}\ }\left\langle {\delta p^{2}}\right\rangle \kappa ^{4} l_{c}^{3}.
\label{k}
\end{equation}
Using Eq. (\ref{Delta_p3d}) and the identification of $l_c$ one can obtain for the scattering length:
\begin{equation} 
l_{s}^{-1}\simeq \frac {\pi ^{1/2}}{16}\ p(1-p)a^{3}\ \left[ \frac {\omega ^{2}}{c^{2}}\frac {\delta \epsilon (\omega ,p({\bf r}))}{\delta p({\bf r})}\right]^{2} .
\label{MFP}
\end{equation}
In the limit of small concentrations, $p\ll 1$, the derivative of $\epsilon(\omega,p)$ with respect to $p$ can be evaluated in different regions of the spectrum.

To assess the value of the scattering length inside the impurity band we pick the center of the defect band $\omega _{c}^{2}=(\omega _{il}^{2}+\omega _{iu}^{2})/2$. For small concentration one obtains
\begin{equation} 
l_{s}^{-1}\left( \omega _{c}\right) \simeq \frac{\pi ^{1/2}}{4}\frac{l^{3}}{l_{0}^{4}\left( \Omega _{1}\right) }\alpha _{1}^{2}\sim \frac{1}{p},
\label{ls_center}
\end{equation}
where 
\[
\alpha _{1} =\frac{\Omega _{1}^{2}-\Omega _{0}^{2}}{\displaystyle{\frac{4\epsilon _{\infty }}{3\left( 2+\epsilon _{\infty }\right) }}d_{0}^{2}+\displaystyle{\frac{2-\epsilon _{\infty }}{2+\epsilon _{\infty }}}\left( \Omega _{1}^{2}-\Omega _{0}^{2}\right)}
\]
does not depend upon the concentration. In the last expression the anticipated small parameter $l/l_0$ has appeared raised to the third power. Taking into account typical values of this parameter for realistic mixed crystals, one can see that this scattering is completely negligible. One may also note that this scattering length increases with concentration, which  has a simple physical explanation -- the greater the concentration the greater the overlap of individual local states, and the closer the system is to a uniform continuous medium.
  
For frequencies from  host bands, the situation is qualitatively different:
\begin{eqnarray}
l_{s}^{-1}\left( \omega \lesssim \Omega ^{(TO)}\right) &\simeq &\left[ \frac{%
\pi ^{1/2}}{16}\alpha _{2}^{2}\left( \frac{\Omega _{0}^{(TO)\;2}}{\omega
^{2}-\Omega ^{(TO)\;2}}\right) ^{4}\right] \frac{pa^{3}}{\lambda ^{4}\left(
\Omega _{0}^{(TO)}\right) }\sim p
\nonumber \\
l_{s}^{-1}\left( \omega \gtrsim \Omega ^{(LO)}\right) &\simeq &\left[
\frac{\pi ^{1/2}}{16}\alpha _{3}^{2}\right] \frac{pa^{3}}{\lambda
^{4}\left( \Omega _{0}^{(LO)}\right) }\sim p
\label{ls_band}
\end{eqnarray}
where 
\[
\alpha _{2} =\epsilon _{\infty }\frac{\frac{\epsilon _{\infty }}{3}d_{0}^{2}\left( \Omega _{0}^{(LO)\;2}-\Omega _{0}^{(TO)\;2}\right) \left( \Omega _{1}^{(TO)\;2}-\Omega _{0}^{(TO)\;2}\right) }{\Omega _{0}^{(TO)\;4}\left( \Omega _{1}^{2}-\Omega _{0}^{(TO)\;2}\right) }
\]
and 
\[
\alpha _{3} =\epsilon _{\infty }\frac{\frac{2}{3}d_{0}^{2}\left( \Omega _{1}^{(LO)\;2}-\Omega _{0}^{(LO)\;2}\right) }{\left( \Omega ^{(LO)\;2}-\Omega ^{(TO)\;2}\right) \left( \Omega _{0}^{(LO)\;2}-\Omega _{1}^{2}\right) }
\]
are constants. This scattering length is much smaller then for impurity polaritons, and is decreasing with the increase of concentration, which is quite a regular behavior.  Indeed, the perfectly ordered at $p=0$ system experiences increasing scattering due to the impurities. This difference between scattering properties of regular and impurity polaritons explains why our approach can be justified to describe properties of the former, while the latter requires more elaborate treatment of disorder.

\section {Discussion}

In this paper we suggested that certain features of optical spectra of mixed crystals in the {\it restrahlen} region
can be  explained if, along with standard optical phonon vibrations, one introduces additional impurity induced 
oscillators with the frequency falling into the {\it restrahlen} of the host crystal. Justification for this 
assumption comes from the concept of local polaritons introduced earlier in Refs.\onlinecite{Rupasov,Classics,Deych,Podolsky} and studies of one-dimensional models,\cite{EuroPhysLett,One-Imp-Transm,Singh,Lyapunov}
which showed how local polaritons develop into an impurity induced band. Local polaritons are significantly different from
local phonons, which were extensively studied in connection with optical properties of mixed crystals. First, they are 
assumed to exist in the {\it restrahlen} region, which in many cases is filled with phonons. Therefore, it is not exactly a 
spectral gap, which is necessary for local states to exist. However,  the dielectric function within the {\it restrahlen} 
remains negative, which means that electromagnetic excitations cannot exist in this frequency region, which is therefore a 
gap for electromagnetic excitations. This situation is reversed compared to the case of pure phonon gaps, which are deprived 
of phonon states, but contain electromagnetic ones. Because of non-zero electromagnetic DOS, local phonons acquire their 
electromagnetic radiation width, and because of non-zero phonon DOS, the local polaritons acquire their phonon 
``radiation'' width. A significant difference between these two situations is that the density of electromagnetic states in 
the phonon gap is so small that the radiative broadening of local phonons is negligible, while the density of phonon states 
in the {\it restrahlen} is large, and local polaritons may or may not survive it. Analyzing known phonon DOS, we found that 
for some crystals like $GaP$, $ZnS$, $CuBr$, $ZnTe$, $CuI$, $SrF_{2}$, $BaF_{2}$, $PbF_{2}$, $UF_{2}$, $CaF_{2}$\cite 
{Phonon-dispersion} there are actually frequency regions with zero phonon DOS, though often quite narrow ones. Therefore, 
the right impurity could in principle give rise to a local state considered in Refs.\onlinecite
{Rupasov,Classics,Deych,Podolsky}. However, {\it restrahlen} region of a much broader class of crystals like $NaCl$, $NaBr$, 
$KCl$, $RbCl$ and many others\cite {Phonon-dispersion}  are filled with LO phonons whose DOS within certain regions is 
relatively small. In the present paper we studied the life-time of local polaritons due to interaction with these LO 
phonons, and found that under regular circumstances this life-time is no shorter than the one due to anharmonicity. We 
argued, therefore, that local polaritons can actually survive interaction with {\it restrahlen} phonon states, and 
contribute to the optical properties of the crystals. Moreover, the presence of some local states within the {\it 
restrahlen} was confirmed experimentally in Ref.\onlinecite{Dean}, where neutral dopants ($S,Sn,Te$) give rise to the local 
electron-phonon state at the frequency slightly below LO of $GaP$. What is interesting is that the region where these states 
reside, has a relatively high density of phonon states, which obviously did not preclude them from existence. All these 
arguments justify the use of local polaritons to describe properties of the {\it restrahlen} of mixed crystals.

The second important property of local polaritons is that their spatial extent is of the order  of optic wavelengths in the 
{\it restrahl}  ($\sim 10^{-3}\ cm$), which is much larger than the size of local phonons (several interatomic distances). 
This fact means that even at residual concentrations of impurities, local polaritons overlap, forming a well developed band. 
We showed in this paper that this band can be described using the continuous medium approximation for the impurity 
subsystem. Neglecting fluctuations of impurity concentration, we derived an effective dielectric function for our model, and 
used it to analyze the  structure of optical spectra of mixed crystals. This dielectric function describes  new impurity 
induced polariton bands, which arise inside the {\it restrahlen} of the host crystal.

The first problem we set out to consider using the concept of impurity polaritons was  weak features in the spectra of so 
called one-mode crystals $Ba/SrF_{2}$, $Ca/SrF_{2}$,\cite {BaSrF2Verleur,BaSrF2Chang} $ZnCdS$, $Mn/ZnTe$,
\cite {Taylor-review} and the one-two-mode behavior of a different group of crystals $In/GaP$,\cite {GaInP} $PbSe/Te$,\cite{PbSeTe} 
$K/RbI$,\cite {KRbIFertel,KRbIRenker} $RbBr/Cl$, $GaAs/Sb$, $InAs/Sb$, $AgBr/Cl$,\cite{Taylor-review} $In/GaAs$.\cite 
{Groenen} The existing descriptions of these types of spectrum\cite{Taylor-review} require a great number of fitting 
parameters. Our model allowed us to explain these types of spectra naturally as manifestations of the impurity polariton 
mode, which reveals itself differently depending upon the relation between fundamental frequencies of crystals at both ends 
of the concentration range, and the frequency of the local polariton. Introducing relaxation in a phenomenological way, we 
considered reflection and transmission spectra, and demonstrated that our model survives rather strong damping, and 
reproduces spectra closed to experimental observations. For rather thin samples our model predicts a resonant enhancement in 
the transmission at the frequency of the impurity polaritons, which accompanies an anti-resonance in reflection. The latter 
was observed in many papers,\cite{Sievers,Taylor-review} but was mostly attributed to the impurity induced absorption. The 
only transmission measurements known to us were performed on pure $CuCl$,\cite{CuClTransm} where transmission was found to 
exhibit the maximum similar to one predicted in our paper. $CuCl$ is a peculiar material, since $Cu$ atoms at low 
temperature can occupy several non-equivalent positions, thereby creating internal defects.\cite{Livescu,OffCenter} These 
off-center $Cu$ atoms can be responsible for the impurity polariton band, and therefore, transmission spectra of $CuCl$ can 
be considered as the first evidence of this band.   

Since we took the retardation into account, we were able consider not only boundaries of the spectra but also dispersion 
laws, DOS, and group velocities of the impurity induced polaritons.  One of the most remarkable properties of these 
excitations is that their group velocity is proportional to the concentration, and can be thousands of times smaller than 
the speed of light in vacuum. The smallness of the group velocity makes dispersion curves of the excitation look almost 
flat. Rather similar dispersion curves were measured experimentally in Ref.\onlinecite{Livescu} in one-two-mode mixed 
crystal $Ga_{0.70}In_{0.30}P$ with the use of Raman spectroscopy. It would be interesting to carry out additional steady 
state and time-resolved experiments in this material, which could verify predictions of our theory and provide more solid 
support for our concept of impurity-induced polaritons. Our approach also allowed us to study scattering of impurity 
polaritons due to fluctuations of concentrations. We found that their scattering length is significantly different from the 
similar characteristics of regular polaritons of the host material. The scattering length of impurity polaritons is 
proportional to the concentration, making the scattering less efficient with an increase of concentration. Quantitatively, 
the scattering length is very large, much larger than the attenuation length due to inelastic damping, making scattering due 
to concentration fluctuations negligible for impurity polaritons. This finding is in agreement with the  results obtained 
for the one-dimensional model,\cite{Lyapunov} and provides a firm foundation for our approach. The scattering length for 
regular host polaritons, at the same time, is inversely proportional to the concentration, and is rather short. Therefore, 
studying optical properties related to these excitations requires more elaborate theoretical approaches used in many papers 
on the subject (see, for instance, Ref.\onlinecite{Bonneville}).

We hope that the present results will revive an interest of experimentalists in the properties of {\it restrahlen} of mixed 
crystals. In our opinion,  it would be interesting to study transmission spectra through thin slabs of one-two mode crystals 
in both sw and time resolved experiments. Such experiments could provide additional insight into properties of impurity 
polaritons, and elucidate their dynamic properties, such as group velocities. Comparing experiment and present theoretical 
results, one can obtain additional information about the material parameters of these systems.  

{\em Note added in proof} A.J. Sievers drew our attention to an alternative explanation of the IR reflection spectra of {\em CuCl}. In Refs. [39], an additional absorption line inside the gap was attributed to the interaction of TO phonons with acoustic phonons arise from by strong anharmonic terms in the potential of {\em Cu} atoms, rather than due to an additional polariton band associated with the off-center ions. Although there exists strong evidence in favor of the model discussed in Refs. \onlinecite{note}, it does not explain, however, the enhanced transmission observed in Ref. \onlinecite{CuClTransm}.

\section {Acknowledgments}
This paper was written in an attempt to answer some questions raised by A.J. Sievers in the course of many discussions. We 
want to express our deepest gratitude to him. We would also like to acknowledge discussions with J.L. Birman and A.A. Maradudin. 
We are indebted to S. Schwarz for reading and commenting on the manuscript. This work was partially supported by NATO 
Linkage Grant N974573 and PSC-CUNY Research Award.



\end{document}